\documentclass[aps,prd,preprint,showpacs,superscriptaddress,amsmath]{revtex4}  
\usepackage{longtable}
\usepackage{graphicx}
\usepackage{amsmath,amssymb}
\usepackage{color}
\usepackage{bbm}


\newcommand{\beq}{\begin{equation}}
\newcommand{\eeq}{\end{equation}}
\newcommand{\bdm}{\begin{displaymath}}
\newcommand{\edm}{\end{displaymath}}

\begin{document}

\title{Characterization of the seismic environment at the Sanford Underground Laboratory, South Dakota}

\author{Jan Harms}
\affiliation{University of Minnesota, 116 Church Street SE, Minneapolis, MN 55455, USA}
\author{Fausto Acernese}
\author{Fabrizio Barone}
\affiliation{Universit\`a degli Studi di Salerno, Fisciano (SA), Italy}
\affiliation{INFN Sezione di Napoli, Napoli, Italy}
\author{Imre Bartos}
\affiliation{Columbia University, New York, NY 10027, USA}
\author{Mark Beker}
\affiliation{Nikhef, National Institute for Subatomic Physics, Science Park 105, 1098 XG Amsterdam, The Netherlands}
\author{J.~F.~J.~van den Brand}
\affiliation{Nikhef, National Institute for Subatomic Physics, Science Park 105, 1098 XG Amsterdam, The Netherlands}
\author{Nelson Christensen}
\affiliation{Carleton College, Northfield, MN 55057, USA}
\author{Michael Coughlin}
\affiliation{Carleton College, Northfield, MN 55057, USA}
\author{Riccardo DeSalvo}
\affiliation{California Institute of Technology, Pasadena, California 91125, USA}
\author{Steven Dorsher}
\affiliation{University of Minnesota, 116 Church Street SE, Minneapolis, MN 55455, USA}
\author{Jaret Heise}
\affiliation{Sanford Underground Laboratory, 630 East Summit Street, Lead, SD 57754, USA}
\author{Shivaraj Kandhasamy}
\affiliation{University of Minnesota, 116 Church Street SE, Minneapolis, MN 55455, USA}
\author{Vuk Mandic}
\affiliation{University of Minnesota, 116 Church Street SE, Minneapolis, MN 55455, USA}
\author{Szabolcs M\'arka}
\affiliation{Columbia University, New York, NY 10027, USA}
\author{Guido M\"uller}
\affiliation{University of Florida, Gainesville, FL 32611, USA}
\author{Luca Naticchioni}
\affiliation{Department of Physics, University of Rome ``Sapienza``, P.le Aldo Moro 2, 00185 Rome, Italy}
\affiliation{INFN Sezione di Roma, 00185 Rome, Italy}
\author{Thomas O'Keefe}
\affiliation{Saint Louis University, 3450 Lindell Blvd., St. Louis, MO 63103, USA}
\author{David S.~Rabeling}
\affiliation{Nikhef, National Institute for Subatomic Physics, Science Park 105, 1098 XG Amsterdam, The Netherlands}
\author{Angelo Sajeva}
\affiliation{Dipartimento di Fisica ``Enrico Fermi``, Universit\`a di Pisa, Largo Bruno Pontecorvo, Pisa, Italy}
\author{Tom Trancynger}
\affiliation{Sanford Underground Laboratory, 630 East Summit Street, Lead, SD 57754, USA}
\author{Vinzenz Wand}
\affiliation{EADS Astrium GmbH, Science Missions \& Systems (AED 41), 88039 Friedrichshafen, Germany}


\date{\today}

\begin{abstract}
An array of seismometers is being developed at the Sanford Underground Laboratory, the former Homestake mine, in South Dakota to study the properties of underground seismic fields and Newtonian noise, and to investigate the possible advantages of constructing a third-generation gravitational-wave detector underground. Seismic data were analyzed to characterize seismic noise and disturbances. External databases were used to identify sources of seismic waves: ocean-wave data to identify sources of oceanic microseisms, and surface wind-speed data to investigate correlations with seismic motion as a function of depth. In addition, sources of events contributing to the spectrum at higher frequencies are characterized by studying the variation of event rates over the course of a day. Long-term observations of spectral variations provide further insight into the nature of seismic sources. Seismic spectra at three different depths are compared, establishing the 4100-ft level as a world-class low seismic-noise environment.
\end{abstract}

\maketitle

\section{Introduction}
\label{sec:Intro}
\begin{figure}[ht!]
\centerline{\includegraphics[width=8cm]{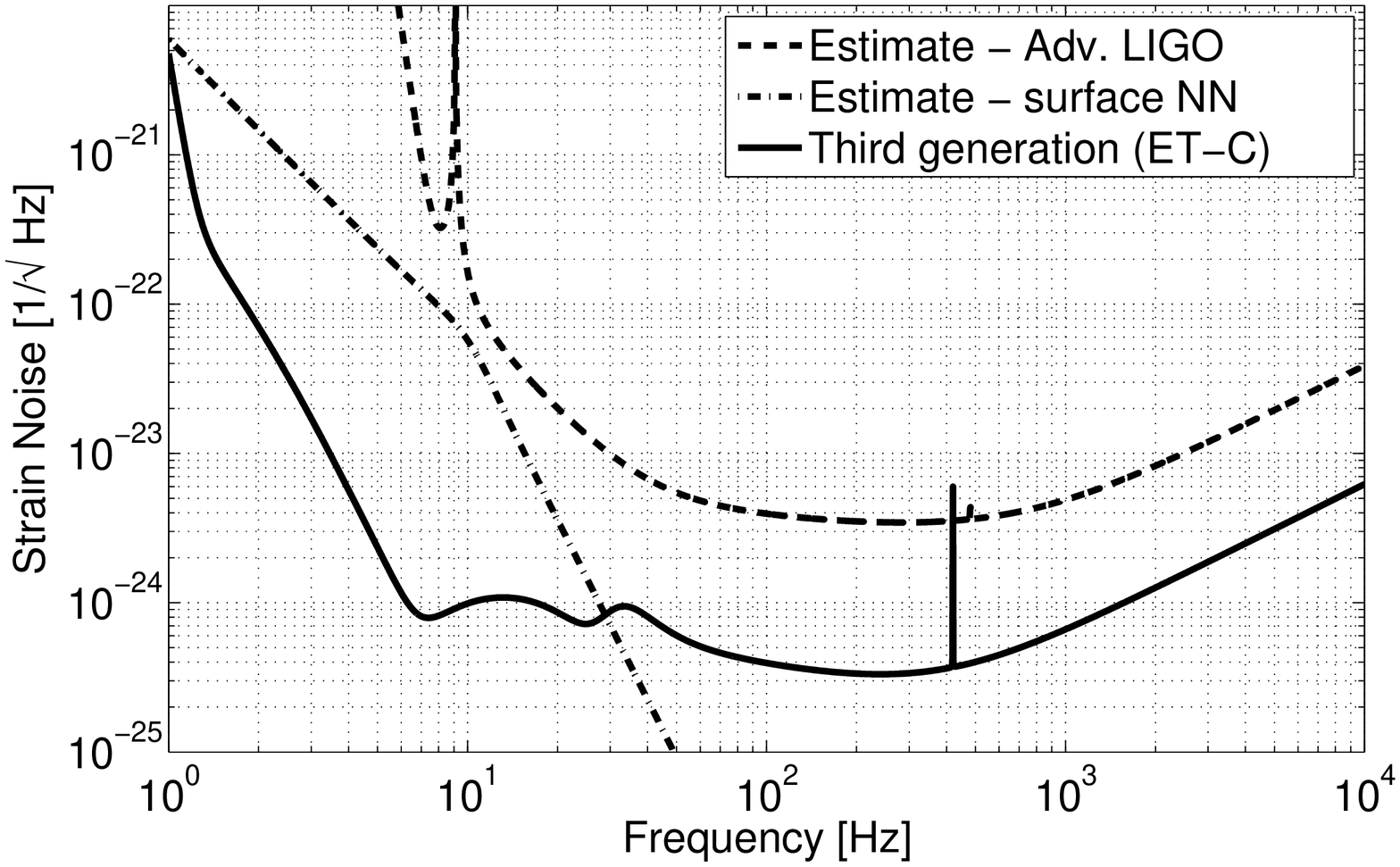}}
\caption{The plot shows an estimate of the sensitivity of the second-generation detector Advanced LIGO \cite{LSC2009a} in comparison with a sensitivity model that was proposed for the third-generation detector ET \cite{HiEA2009}. The Newtonian-noise curve is based on a model that approximates a typical spectrum of seismic surface waves measured at the Hanford observatory of the LIGO detectors, and is based on the characteristics of local geology \cite{HuTh1998}.}
\label{fig:adLIGO}
\end{figure}

Seismic waves produce perturbations of the gravity field, which are predicted to cause detectable levels of Newtonian noise (NN) or gravity-gradient noise in future gravitational-wave (GW) detectors \cite{Sau1984,BeEA1998,HuTh1998,Cre2008}. Whereas the sensitivity of currently operating detectors is not limited by NN \cite{LSC2009b,LuEA2006,VIR2008,Tat2008}, second and third-generation detectors will be sensitive to gravity perturbations at 30\,Hz and below.  Figure \ref{fig:adLIGO} shows estimates for the sensitivity of second and third-generation detectors together with an estimate of Newtonian noise at a present surface site.

Third-generation GW detectors will be designed with enhanced sensitivity below 10\,Hz based on improved suspension systems and optimized material properties to mitigate the seismic and thermal noise \cite{RoEA2002,RobEA2004}. Moreover, quantum-non-demolition techniques are being investigated to cancel part of the optical quantum noise \cite{Che2003,KLMTV2001,HaEA2003}. This leaves the question whether the NN, which is directly imprinted on the test-mass motion without the possibility to shield against it, can be mitigated. One obvious improvement would be to identify a detector site with a comparatively low level of seismic and NN noise, which also includes the possibility to construct the detector under ground. Underground seismic noise at depths of about 1\,km can be an order of magnitude weaker than surface noise above 1\,Hz \cite{Bor2002,CarEA1991,Dou1964}, but further NN mitigation by two orders of magnitude is required to achieve good sensitivities at frequencies close to 1\,Hz, as shown in Figure \ref{fig:adLIGO}. 

Choosing a quiet underground location only gains a limited amount of frequency coverage. A further step to push the sensitivity threshold at even lower frequency is to subtract from the GW data an estimate of NN based on seismic measurements. The idea is to combine the data of several seismometers to define an adaptive filter that minimizes the variance of the GW data. It has been shown that this problem has a trivial solution under ideal conditions, i.e.~homogeneous rock, distant sources, and negligible surface effects. In this case, only a few seismometers would be required for the NN subtraction \cite{HaEA2009b}. The main experimental challenge will be to understand how much realistic seismic environments differ from ideal conditions, how many seismometers are actually necessary to reach a given GW sensitivity target, and where the seismometers should be located.  

A small underground array of broad-band seismometers has been constructed to study the seismic environment of the former Homestake mine in the Black Hills of South Dakota \cite{CaEA1991}. In this paper, we present a characterization of local seismic noise and seismic disturbances. Disturbances include mundane sources like water pumps and the ventilation system, and short events that are mainly associated with excavation blasts and subsequent rock fall, or rail traffic nearby seismic stations. Based on seismic data recorded in December 2009 and January 2010, we will show that local seismic events contributing to frequencies below 30\,Hz occur rarely at Homestake. Besides the before-mentioned identified disturbances, these signals mainly originate from unidentified sources (e.g.~natural rock fracturing and surface sources). 

In Section \ref{sec:Array}, we describe the configuration of the seismic array, and discuss errors with respect to timing, seismometer positions and instrumental noise. In Section \ref{sec:Blasts}, we present our analysis of seismic events. Daily event rates averaged over two months are presented, and first conclusions are drawn concerning the nature of the sources. In Section \ref{sec:Continuous}, results are shown based on a continuous long-term analysis of seismic data. A more detailed examination of the oceanic microseisms is presented in Section \ref{sec:Secondary}. Our conclusions are presented in Section \ref{sec:Conclude}.

\section{Experimental setup}
\label{sec:Array}
\begin{figure}[t]
\centerline{\includegraphics[width=8cm]{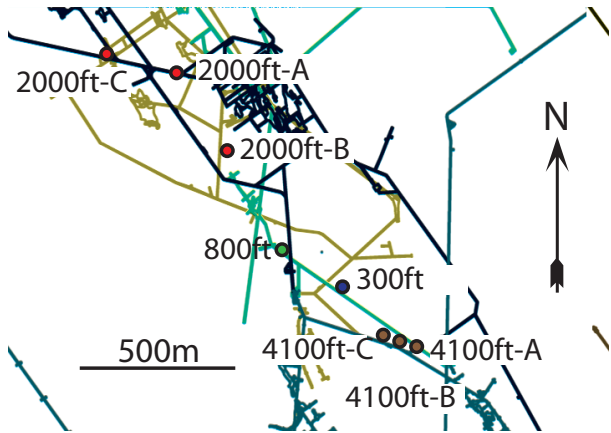}}
\caption{The figure shows a small section of the levels used for the seismic array. The circles indicate the locations of seismometers. The 300\,ft level is shown in blue, the 800\,ft in green, the 2000\,ft level in red, and the 4100\,ft level in brown.}
\label{fig:Stations}
\end{figure}

The underground array consists of 8 environmentally shielded and isolated seismic stations at 4 different depths. One Streckeisen STS-2 seismometer is located at the 4100\,ft level, two G\"uralp CMG-40T at 300\,ft and 4100\,ft, and five Nanometrics T240: one at 800\,ft, three at 2000\,ft and one at 4100\,ft depth. The configuration of the array is shown in Figure \ref{fig:Stations}. The readout system includes 18-bit ADCs and preamplifiers with a gain of 100 with the focus on small noise levels, while allowing seismic events uninteresting to us to saturate in our readout chain. Together with the seismometers, most stations are equipped with environmental sensors including magnetometers, humidity sensors, thermometers and barometers.

All seismic stations are connected via optical fibers to a computer at the surface, which stores the data and serves as master clock for the network-time protocol (NTP) synchronization of underground stations. The surface computer is NTP synchronized with public web NTP servers. The dominant absolute timing error comes from the LabVIEW based real-time processing of the data acquisition at each station. This error is mostly eliminated by linear-regression of time stamps in data files over an entire day. Although certain systematic timing errors cannot be identified or corrected by this method, correlation measurements between different stations show that timing errors are consistent with the calculated regression errors, which are typically close to 2\,ms per second and do not exceed 5\,ms per second (unless network connectivity is temporarily lost, which did not occur in December 2009 and January 2010). 

Measures were taken at all stations to optimize the seismometer response to seismic fields. First, seismometers were installed either on existing concrete platforms that are known  (by sounding and physical edge exploration) to be solidly connected to the rock, or loose and waste rock was removed before a new concrete platform was poured directly onto the bedrock. Second, placing seismometers on granite tiles grouted to the concrete improved seismic spectra considerably above 10\,Hz \cite{AcEA2008}. We have not yet plastered the seismometer feet to the granite tiles as is recommended in \cite{PaVe1994} to further improve the high-frequency spectra. Third, a multi-layer isolation frame of rigid thermal and acoustic insulation panels was built around each seismometer to further stabilize the thermal environment and to achieve suppression of acoustical signals and air currents. However, we found that insulation panels cannot guarantee a quiet environment. For example, the noise spectra of seismic stations at the 4100\,ft level depend significantly on location. Two stations are located in concrete rooms at the side wall of a drift with weak but noticeable air currents. There seismic-noise spectra are about a factor 10 stronger in amplitude than spectra from the third 4100\,ft station that is located close-by in the same drift, but off the main air flow inside a small and well isolated chamber that used to serve as storage room. For this reason, data from only one station at the 4100\,ft were further analyzed. As we will see in the next section, the seismic spectrum of this station proves that the 4100\,ft level provides a superb low seismic-noise environment suitable for our studies. 

Table \ref{tab:Stations} lists some properties of the seismic stations. Based on our experience with station designs, we assigned a quality measure to each station. Q1 signifies a station that needs substantial redesign and/or relocation to become a valuable contribution to the array, noise spectra above 3\,Hz of a Q2 station are expected to change significantly if their readout system and station design are improved, spectra of Q3 stations are expected to change significantly if the readout system is improved, and Q4 signifies a station where only minor changes of the seismic spectrum are expected if the station design or readout system are further improved. The problems with Q1 and Q2 stations would be resolved if the seismometers were moved to more favorable locations, e.g.~to properly sealed off blind drifts, small chambers off the drift, or boreholes. The rating is based on the quality of spectra above 3\,Hz. From coherence measurements, we can conclude that designs and readout systems of all stations are adequate to provide high-quality data within the pass-band of the seismometers up to 3\,Hz. Seismometer positions were determined from mine maps. Position errors are smaller than 2\,m.
\begin{table}[ht!]
\begin{tabular}{|l|c|c|p{2.5cm}|}
\hline
Station & Seismometer & Position (E,N) [m] & Characteristics \\
\hline\hline
300\,ft & CMG-40T & (71,21) & horizontal access, calibrated, Q3\\
\hline
800\,ft & T240 & (-88,124) & inoperative, calibrated, Q3\\
\hline
2000\,ft A & T240 & (-378,598) & calibrated, Q3\\
\hline
2000\,ft B & T240 & (-234,380) & calibrated, Q2\\
\hline
2000\,ft C & T240 & (-523,586) & Q3\\
\hline
4100\,ft A & STS-2 & (347,-155) & calibrated, Q4\\
\hline
4100\,ft B & T240 & (274,-132) & inoperative, Q1\\
\hline
4100\,ft C & CMG-40T & (187,-104) & inoperative, Q1\\
\hline
\end{tabular}
\caption{The horizontal position is given along cardinal directions relative to the Yates shaft of the Homestake mine, which is currently the main access to underground levels. Characteristics are a subjective collection of properties that appear noteworthy in the context of this paper. This table refers to the setup during December 2009 and January 2010, since the array is subject to improvements and changes. Stations termed ``inoperative`` did not acquire data during these two months. Note that data analyzed for this paper were recorded by instruments whose response (relative to other instruments) was measured. These instruments are characterized as ``calibrated``. Well before December 2009, seismic spectra were further validated by side-by-side measurements at some of the stations.}
\label{tab:Stations}
\end{table}

\section{Analysis of seismic events}
\label{sec:Blasts}
In this section, we investigate daily event rates calculated from 2 months of data.  To search for large \emph{events} in the seismometer data we used a wavelet based tool called \emph{KleineWelle} \cite{ChEA2004}. KleineWelle was originally designed to find excess power signals, specifically gravitational-wave burst type events (such as those from supernovae), in the data from interferometric gravitational wave detectors. It proved an effective method for finding large amplitude and short time-scale transients in the data. Therefore, it has been applied as a means to find spurious noise sources; the Laser Interferometer Gravitational Wave Observatory (LIGO) uses KleineWelle to find noise \emph{glitches} in interferometer auxiliary channels as a means of identifying the origin of noise events seen in the interferometer output \cite{Bla2008}. 

\begin{figure}[t]
\includegraphics[width=8cm]{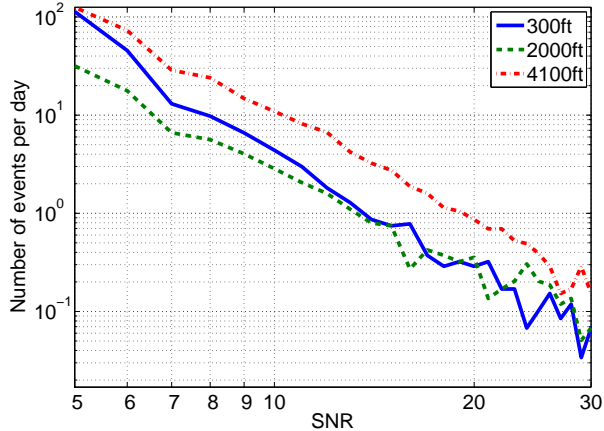}
\caption{The plot shows the number of identified events per day as a function of SNR. The numbers are highest at the 4100\,ft level because the seismic noise is lowest. Taking the elevated seismic-noise spectrum at the 300\,ft level into account, event rates are substantially higher at the 300\,ft than at the other two levels.}
\label{fig:EventSNR}
\end{figure}

All events are included that have signal-to-noise ratio $\rm SNR\geq4.9$. It follows that the total number of identified events also depends on the spectrum of the ambient seismic noise. Results are presented using data from the 4100\,ft-A seismometer, which measures the quietest ambient seismic-noise spectrum. The plot in Figure \ref{fig:EventSNR} displays the total number of identified events per day as a function of SNR obtained by averaging daily event numbers over two months. For each integer SNR $\sigma_0$, the number of events is determined by collecting all events whose SNR $\sigma$ obeys $-0.5<\sigma-\sigma_0<0.5$. The curves can be approximated by $r(\sigma) \propto \sigma^{-4}$. 
Taking the seismic-noise spectra into account, event rates at the 2000\,ft and 4100\,ft levels are very similar. By this we mean to multiply the 2000\,ft rates in Figure \ref{fig:EventSNR} by the averaged ratio of seismic-noise spectra above 1\,Hz (2000\,ft over 4100\,ft, see Figure \ref{fig:SpecComp}). In comparison, noise-corrected event rates of the 300\,ft level are significantly higher than at the other two levels. The total daily numbers of events with $\rm SNR\geq 4.9$ are 200, 80, and 300 at the 300\,ft, 2000\,ft, and 4100\,ft levels respectively. In the future, it will be important to locate the sources, and to find out whether the events are anthropogenic or from natural sources. For example, whether seismic waves originate from the surface or from deep underground, or whether the wave propagates horizontally or vertically, has important implications for NN filtering.
\begin{figure}[t]
\includegraphics[width=8cm]{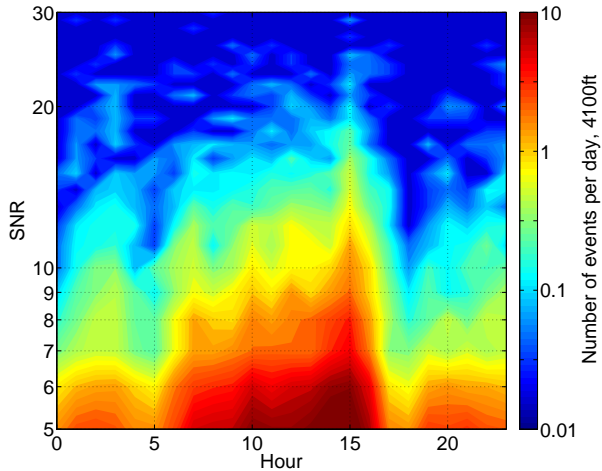}
\caption{Event rates depend on the hour of the day. During the main working hours between 6 AM and 4 PM local time, event rates at the 4100\,ft level are higher by an order of magnitude compared to the rates during nights. The diurnal variation of event rates is strongest for small SNRs indicating their anthropogenic origin.}
\label{fig:EventCont}
\end{figure}
First insight into this problem can be gained by observing the diurnal variation of the rates. As shown in Figure \ref{fig:EventCont} for the 4100\,ft A station, event rates are highest between 6 AM and 4 PM (local time) and constantly low during the night. Also, variations are higher for low-SNR events than for high-SNR events, which suggests that low-SNR events are typically anthropogenic, whereas the majority of high-SNR events have natural sources. Here, anthropogenic events also include anthropogenically triggered natural events like rock fracturing as a response to changes in stress (caused by pumping or blasting) as described in \cite{BMJ2009}.

To relate event rates to a prediction of the available quiet time during the operation of a GW detector, one needs to know the performance of the filter that subtracts the NN. If the transfer function between ground motion and test-mass displacement can be estimated with high accuracy, then in principle the gravity perturbation even from high-magnitude events can be subtracted guaranteeing long quiet times. This is a more serious problem in subtracting seismic disturbances transferred through the suspension system, which is a complex system whose transfer function cannot be estimated accurately due to the instrumental noise of the seismometers or the inherent noise of the suspension. Also, non-linear effects may be relevant when describing the impact of strong seismic disturbances on the suspension system. In conclusion, the definition of quiet times will follow from the studies of filters and how well they perform as a function of event magnitude.

We also searched for coincident events between seismometers and a single high-sensitivity magnetometer that was installed at the 800\,ft level to study a possible relation between seismic and magnetic signals. We found no evidence for magnetometer events in coincidence with seismic signals, but it should be noted that the 800\,ft seismometer did not produce data during this study, so the analysis only involved seismometers at greater distance to the magnetometer. The search needs to be repeated with the magnetometer placed at an operating seismic station.

Initially, we planned to use observations of blast waves from a known excavation site to estimate seismic speeds. The idea failed as will be explained in the following, but the investigations led to other important conclusions. Blast waves originate approximately from a line between the points (533\,m, 566\,m) and (568\,m, 752\,m) at the 4850\,ft level (same coordinate convention is used as in Table \ref{tab:Stations}). Since blasting times are only known approximately, the measurement of arrival times of the wave at the seismometers needs to be triggered by one station. Subsequently, the differential propagation distances to other seismometers can be divided by differential times-of-arrivals to estimate the speed of seismic pressure waves. The differential distances between seismometers and excavation site are listed in Table \ref{tab:Distances} using the 4100\,ft A seismometer location as reference point.
\begin{table}[ht!]
\begin{tabular}{|l|c|c|c|c|}
\hline
Station & 300\,ft & 800\,ft & 2000\,ft A & 2000\,ft B \\
\hline
Distance & 754\,m & 643\,m & 438\,m & 361\,m \\
\hline
Station & 2000\,ft C & 4100\,ft A & 4100\,ft B & 4100\,ft C \\
\hline
Distance & 547\,m & 0\,m & -1\,m & 7\,m\\
\hline
\end{tabular}
\caption{The table lists the additional distances a blast wave from the excavation zone needs to propagate to reach the seismometers once it has arrived at the 4100\,ft A station. Here, we assume that the wave is spherically symmetric.}
\label{tab:Distances}
\end{table}
For example, when the blast wave arrives at the 4100\,ft A seismometer, it needs to travel an additional distance of 361\,m to reach the 2000\,ft B station. The following problems made a travel-time analysis difficult:
\begin{itemize}
\item The sampling frequency of 128\,Hz used at that time was still too low to accurately resolve the first arrival of a blast wave. Due to the short differential travel distances, timing errors of one sampling period are substantial.
\item The seismic noise background at the most distant seismometer at the 300\,ft level was too strong to identify the relatively weak first arrival at this station.
\item The observed peak amplitudes of certain events were significantly smaller at the 2000\,ft A than at the nearby 2000\,ft B seismometer. The 2000\,ft A station lies within a structurally complex region that features a transition between three rock formations. It is conceivable that high-frequency wave components are reflected efficiently (low-frequency components are not affected by small scale structures as illustrated by the fact that the microseismic peaks are measured with equal amplitudes at all stations). It is also possible that despite all efforts the old concrete platform that supports that station has a bad coupling to the hard rock.
\end{itemize}
The attempted blast analysis has therefore demonstrated that the readout system needs to be improved. In particular, higher sampling frequency and better synchronization of seismometers is required. First steps have already been taken to install a timing-distribution system that provides sub-microseconds timing accuracy for all stations. It is based on optical-fiber communication and compensates for light-travel times between different units of the timing system \cite{BaEA2008}.

\section{Continuous long-term observations}
\label{sec:Continuous} 
A continuous long-term study of seismic spectra provides important clues into the nature of seismic noise and regular seismic disturbances. While our array has been monitoring the seismic noise at Sanford Lab for over a year, in this section we present results based on two months of data acquired in December 2009 and January 2010, which have now reasonably good and consistent sensitivity level. This period of time was chosen because operation of seismic stations was less stable before December 2009, making a continuous long-term study difficult. Seismic spectra are calculated using 128\,s contiguous blocks of data sampled at 128\,Hz. The sampling frequency and duration are chosen in accordance with the corner frequencies of the seismometers' pass-bands (common range of pass-band for the three seismometer models is approximately 40\,mHz to 40\,Hz). The basic approach here is to observe the spectral densities at specific frequencies over long periods of time, and to use these time series at each frequency for further analyses. Results are qualitatively different near the surface and deep underground. Therefore, we will compare results from two different levels, 300\,ft, and 4100\,ft, the 300\,ft seismometer being closest to the surface.
\begin{figure}[t]
\includegraphics[width=8cm]{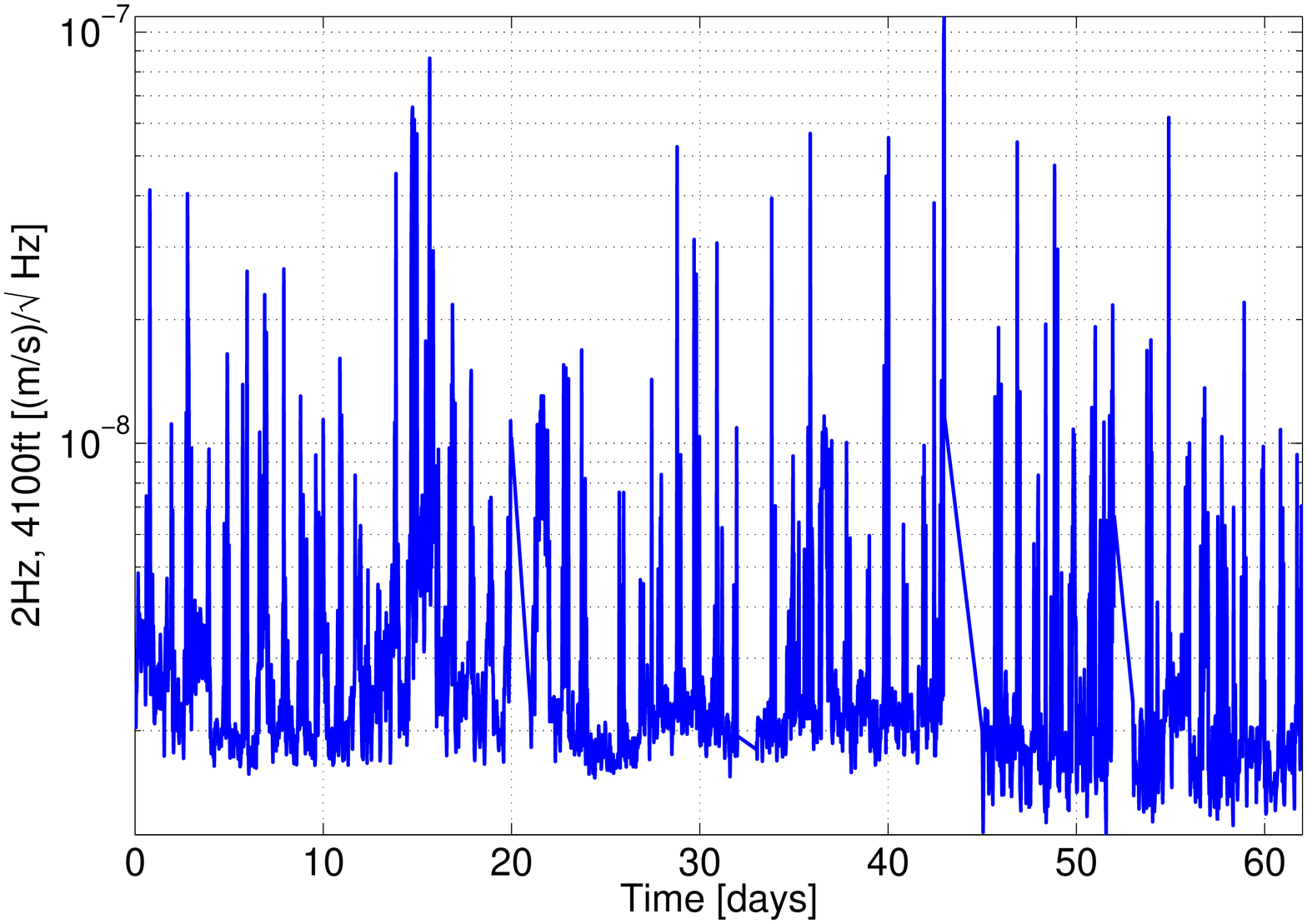}\\
\includegraphics[width=8cm]{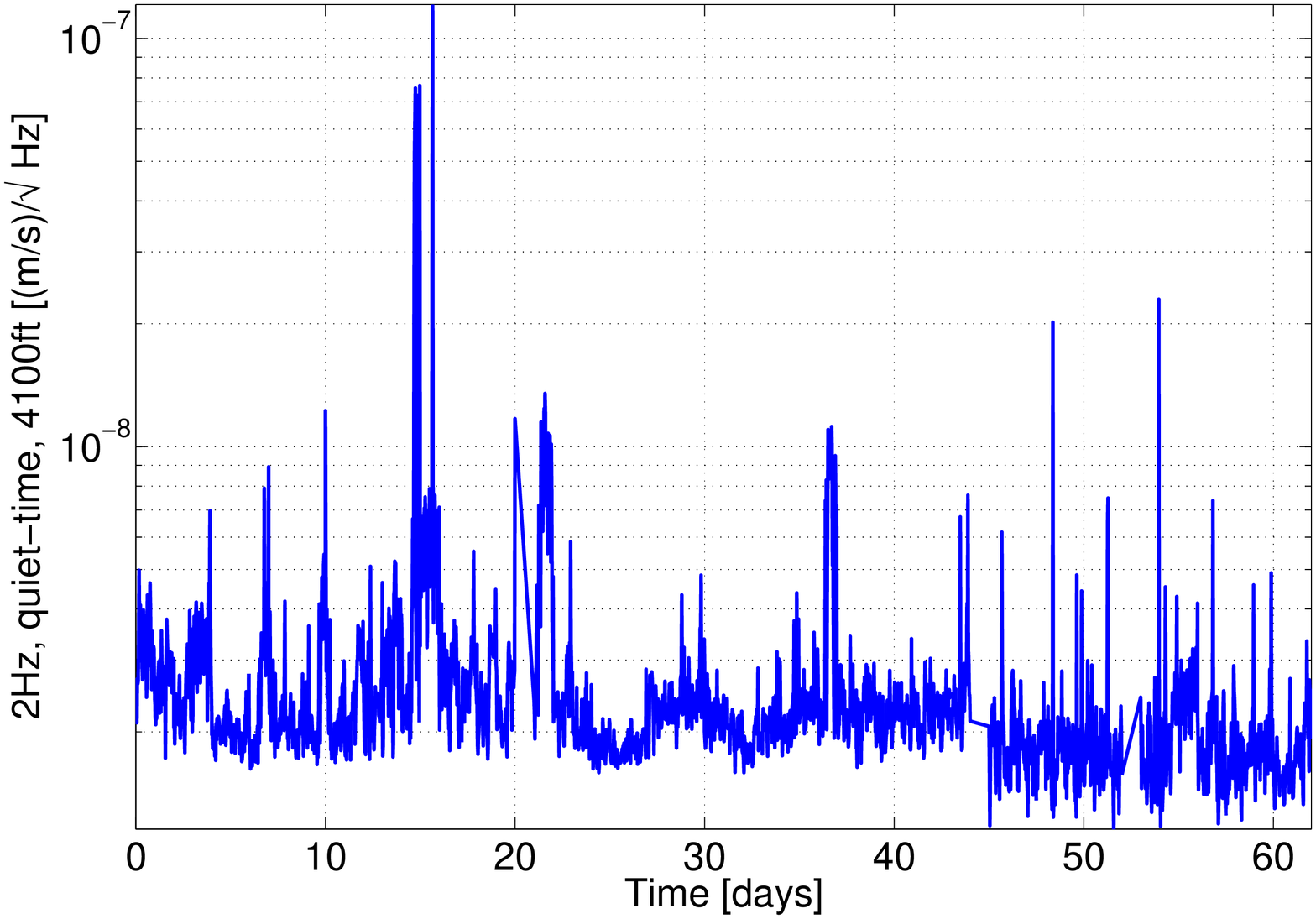}
\caption{The upper plot shows the root spectral density at 2\,Hz measured at the 4100\,ft level over 62 days in December 2009 and January 2010. Each point is a half-hour average of the spectral density calculated from 128\,s stretches. The upper plot contains all disturbances (the strongest disturbances being the excavation blasts), whereas the lower plot only contains longer-lasting disturbances that are not removed by the cleaning algorithm. As can be seen in both plots, the 2\,Hz amplitude decreases slightly after about 45 days and a 2-day interruption of the data acquisition. This change is caused by a change in corner frequency of the low-pass filter in mid January suppressing aliasing of high-frequency noise into the seismic spectrum.}
\label{fig:Longtime}
\end{figure}

We are mainly interested in characterizing the seismic noise level in quiet times. Since daily excavation blasts at the 4850\,ft level and individual seismic events have a great impact on these results, we apply a cleaning algorithm that removes short-duration transients and identifies quiet-time data. It was applied to the data from all of our seismometers. Among the events of significant amplitude ($\rm SNR\geq 4.9$) as identified by KleineWelle, we searched for events that were coincidently seen by KleineWelle in seismometers at different locations; various coincidence time windows (0.1\,s, 0.5\,s, 1.0\,s) were used. We defined quiet times for our analysis by removing periods when there were significant events seen coincidently by KleineWelle in seismometers at two or more locations. 
\begin{figure}[t]
\includegraphics[width=8cm]{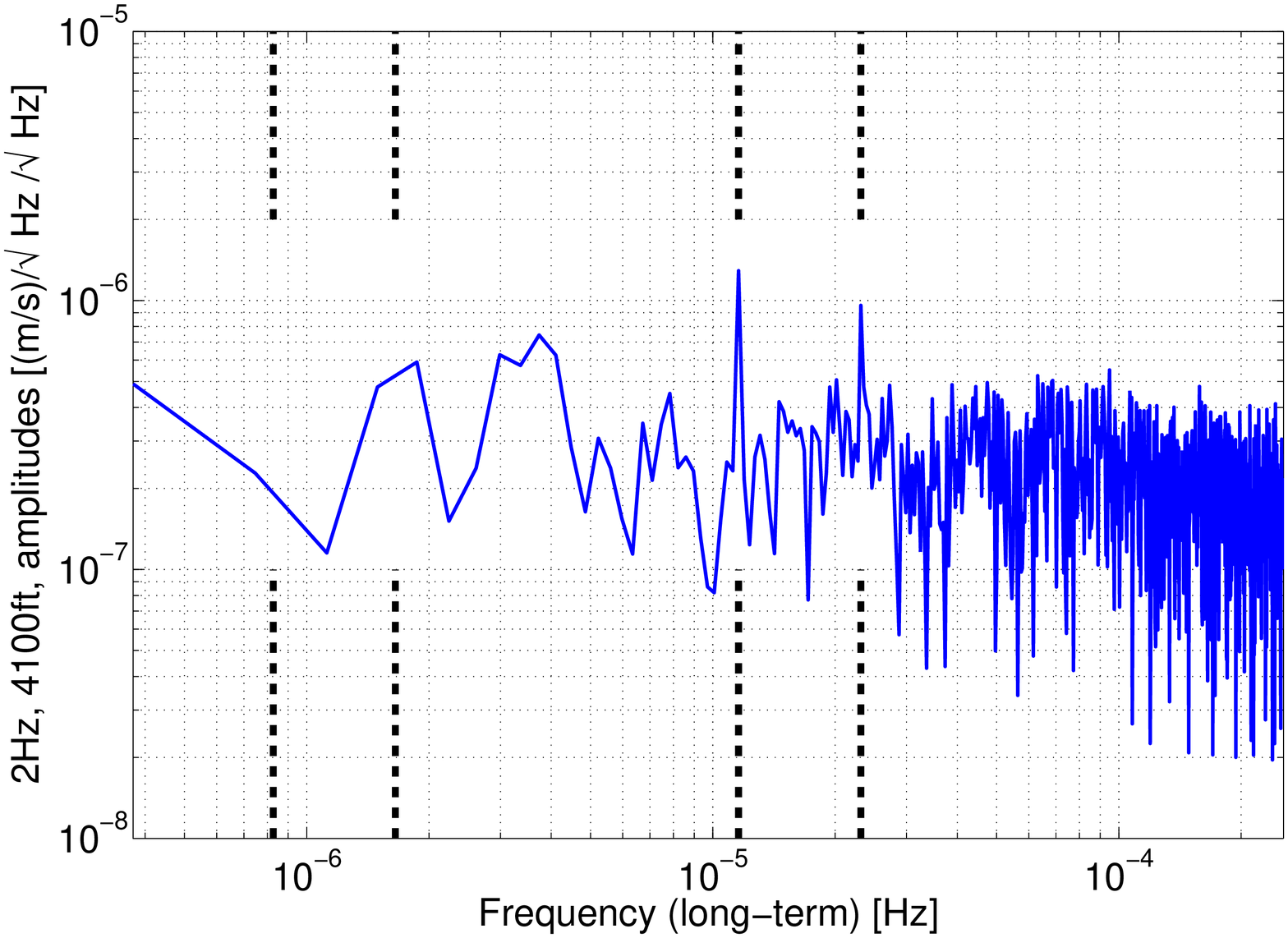}\\
\includegraphics[width=8cm]{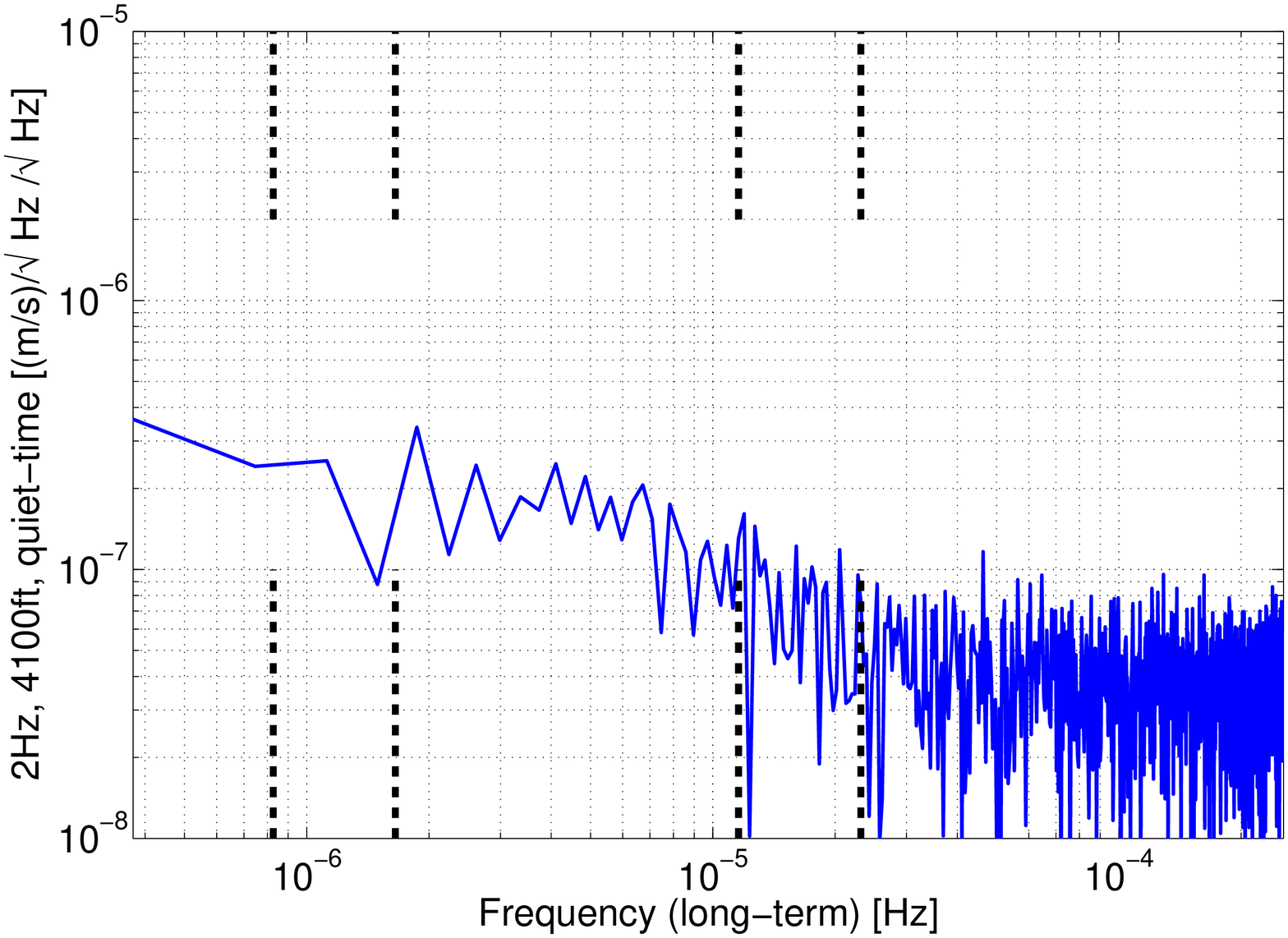}
\caption{The two plots display the root spectral densities of the signals shown in Figure \ref{fig:Longtime}. The 4 vertical dashed lines designate 14\,day, 7\,day, 1\,day and 0.5\,day periods. The peaks at 0.5\,day and 1\,day period in the upper plot are caused by daily blasting, and since blasting follows a weekly cycle (no blasting on weekends), the 7\,day peak is also stronger in the upper plot than the lower quiet-time plot.}
\label{fig:Spec2Hz}
\end{figure}
The 128\,s stretches that contained one or more disturbances were then omitted from further analyses. The times of the observed large signals in the seismometers also provided us with the ability to easily identify and examine individual interesting events (natural or anthropogenic). Figure \ref{fig:Longtime} shows a comparison of spectral evolution at 2\,Hz before and after the cleaning algorithm is applied. A few disturbances remain since the algorithm only removes events that show excess power relative to a background measured within a limited time period around the event. Whereas local excavation blasts are all removed in this way, drilling operations that can last up to a few hours survive the cleaning process. Figure \ref{fig:Spec2Hz} shows the root spectral densities of the time series shown in Figure \ref{fig:Longtime}. The spectrum in the upper plot is based on the uncleaned data. It exhibits strong peaks at 0.5\,day and 1\,day period, and a less pronounced peak at 7\,day period that does not appear in the quiet-time spectrum.

\begin{figure}[t]
\includegraphics[width=8cm]{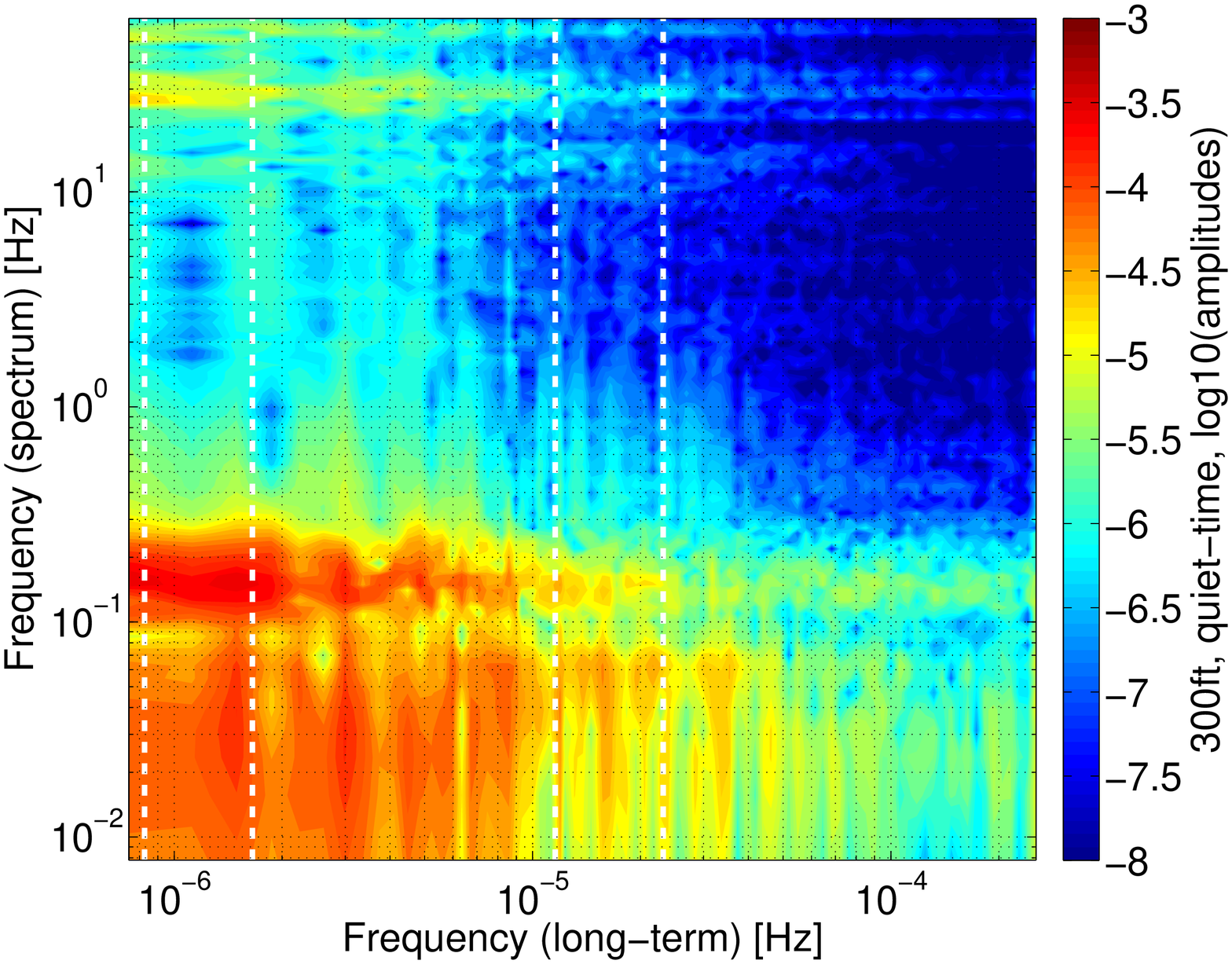}\\
\includegraphics[width=8cm]{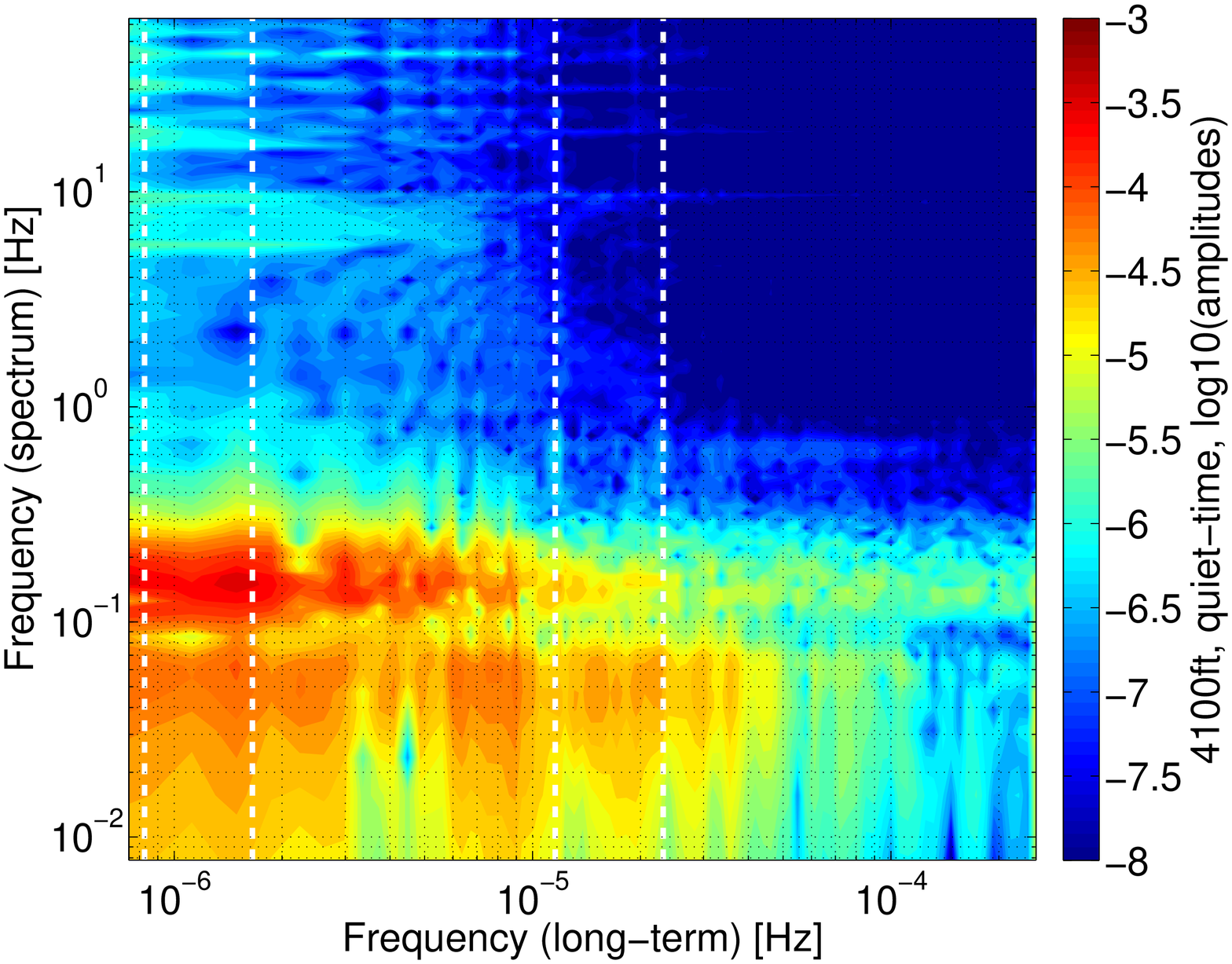}
\caption{The contour plots show the amplitudes that describe the long-term evolution of seismic spectra at 300\,ft and 4100\,ft. The 4 vertical white, dashed lines designate a 14\,day, 7\,day, 1\,day and 0.5\,day period. Spectral variation is generally stronger closer to the surface. At frequencies smaller than 0.1\,Hz, the evolution of seismic amplitudes at 300\,ft is strongly correlated to surface wind speeds, whereas the corresponding amplitudes at 4100\,ft are uncorrelated with wind speeds (see Figure \ref{fig:Wind}). Sources of spectral variability at 4100\,ft at low frequencies include earthquakes and changes in the primary microseismic peak. At frequencies above 0.3\,Hz, seismic amplitudes at 300\,ft show additional variations due to anthropogenic noise, water pipes and surface waves from ventilation fans that are located a few hundred meters away from the station.}
\label{fig:Specdrift}
\end{figure}
All following results are based on quiet-time data. Calculating the amplitudes of the long-term evolution for each frequency bin, all these spectra can be combined into a contour plot as shown in Figure \ref{fig:Specdrift}. Narrow-band features at higher frequencies are suppressed in the contour plots since long-term amplitudes are averaged over frequency intervals that have equal lengths on a logarithmic scale. This leads to averaging over a greater number of bins at high frequencies. At first, we point out the similarities of contour plots calculated for different stations. Coherence of the secondary microseismic peak measured at 300\,ft and 4100\,ft is close to maximum, and therefore variations of amplitudes between 0.1\,Hz--0.3\,Hz are nearly identical at both stations. Amplitudes vary weakly at these frequencies on time-scales shorter than a day. There is a general trend that long-term variation of seismic amplitudes is weaker on shorter time scales. It is difficult to interpret this result, but it is ultimately related to the stability of sources of seismic noise. One should keep in mind that long-term variations at frequencies below 10\,mHz are underestimated since the mean value of time series is subtracted before calculating the FFT of 128\,s data stretches. 

\begin{figure}[t]
\includegraphics[width=8cm]{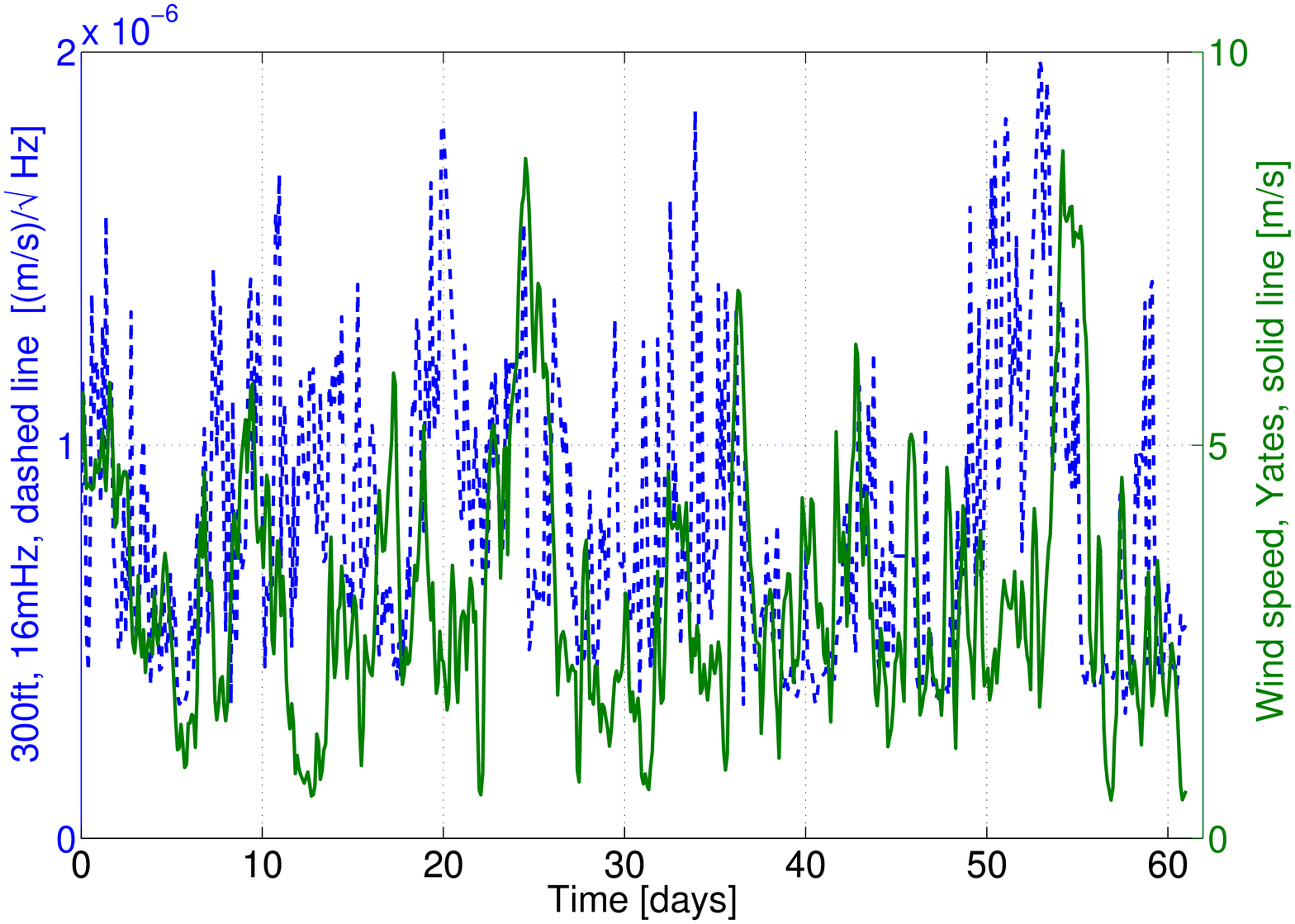}
\caption{The dashed curve shows the root spectral density of ground velocity in vertical direction measured at the 300\,ft level and at 16\,mHz. The solid curve shows the wind speed measured each minute on the roof of one of the mine buildings. To calculate coherence, the wind-speed data were resampled to a half-hour sampling to multiply them with the seismic data. The coherence between the plotted time series is 0.16 at this level, whereas the coherence between wind speeds and ground motion at deeper levels is less than 0.02.}
\label{fig:Wind}
\end{figure}

Concerning the difference between variations at the 300\,ft and 4100\,ft levels, we first observe that variations are generally stronger at the 300\,ft level. For example, the primary microseismic peak that contributes to frequencies between 30\,mHz -- 70\,mHz is concealed at 300\,ft by wind-generated ground motion, which was confirmed measuring coherence between wind speeds and seismic amplitudes over two months. Wind is a major source of microseisms near the surface \cite{WiEA1996}. Figure \ref{fig:Wind} shows a plot of ground velocity measured at the 300\,ft level and surface wind speeds. 
\begin{figure*}[t]
\includegraphics[width=\textwidth]{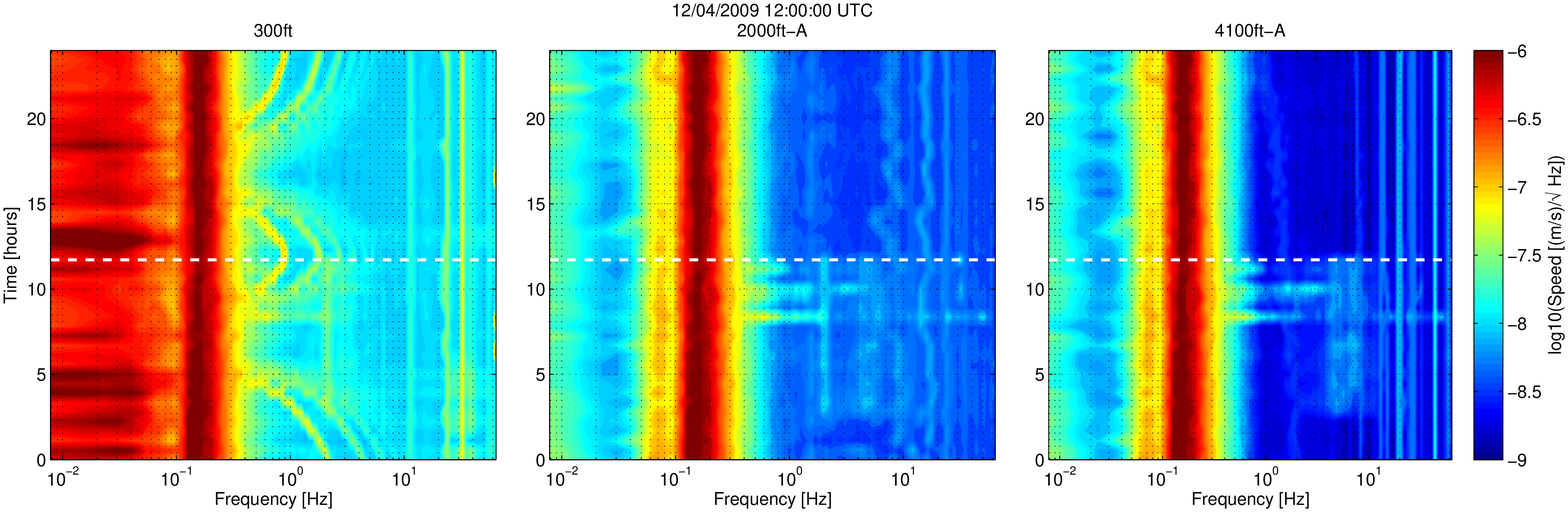}
\caption{The figure shows 24h time-frequency plots for three different levels. The dashed horizontal line indicates UTC 00:00 (5pm local time). December 4th was a Friday, which is typically chosen for ventilation-fan maintenance (ventilation lines at 300\,ft change in frequency). Working activities usually stop before 4:30pm, which is consistent with the spectra at 2000\,ft and 4100\,ft. High-frequency lines at the 4100\,ft disappeared after the corner frequency of the low-pass filter had been optimized mid January 2010. There was no excavation blast on December 4th.}
\label{fig:Snapshot}
\end{figure*}
The coherence between the two time series, 0.16, is small since earthquakes and anthropogenic noise still dominate the low-frequency spectrum. However, the evolution of the time series is clearly related over briefer periods of time, and coherence of surface wind with seismic noise at 2000\,ft or 4100\,ft and at 16\,mHz is significantly smaller (0.01 and -0.02 respectively) lying within the estimation error of coherence. It should be clear that the lack of correlation between underground seismicity at 2000\,ft depth and surface wind speeds cannot be explained by the evanescent character of fundamental Rayleigh waves in homogeneous rock since the wavelength $\lambda_{\rm R}$ at these low frequencies is hundreds of kilometers. One possible mechanism is that these surface waves are generated inside a low-speed surface layer with much smaller wavelength $\lambda_{\rm surf}$. All spatial modes with $\lambda_{\rm surf}\leq\lambda_{\rm R}$ are evanescent surface modes with decay length $\lambda_{\rm surf}$. Small-scale features of the surface displacement field would be further amplified if the mean distance of uncorrelated sources at the surface was much smaller than $\lambda_{\rm R}$. We do not have an understanding of these sources yet, but one likely mechanism is that wind generates turbulences around buildings, trees, and complex surface profiles, which would act back on these structures and lead to a very dense pattern of uncorrelated seismic sources that are usually not directly connected to the hard rock. The long-term evolution of the strength of these turbulences would follow mean wind speeds. It will be important to study this mechanism in more detail, and to include other surface sources like pressure fluctuations and specific anthropogenic disturbances.

Increased variations at frequencies above 20\,Hz at the 300\,ft level are of unknown origin, but it seems likely that they are related to an electronic disturbance since the variations are associated with narrow-band lines that cannot be ventilation lines. Ventilation fans are driven at different frequencies, according to the required load, and switched off regularly for inspection purposes, which can be followed in time-frequency plots. Also, the fundamental frequency of ventilation lines is usually smaller than 10\,Hz (exceeding 10\,Hz only briefly a few times a month), and higher harmonics (observed up to 6th order) are much weaker than the lines observed above 20\,Hz. Only the fundamental frequency of ventilation lines is found in spectra at the 4100\,ft, and its amplitude is weak (less than a factor 3 above seismic noise). Anthropogenic disturbances are present at all levels below 10\,Hz. Excavation blasts and subsequent rock fall contribute to all frequencies of the calculated spectra. An example of 24h time-frequency plots at three different levels can be seen in Figure \ref{fig:Snapshot}. It features a ventilation inspection that can be followed in the 300\,ft plot, and working activities inside the mine that contribute to excess noise at the two deeper levels. The secondary microseismic peak between 0.1\,Hz and 0.3\,Hz was comparatively strong on December 4th. We will investigate its magnitude and frequency evolution in more detail in the next section. 

The question that we attempt to answer next is whether the spectral variations of seismic noise at different frequencies are correlated. Since we are not yet interested in characterizing anthropogenic sources, the analysis will focus on the January 2010 data acquired at the 4100\,ft. We do not consider the December data since seismic noise from frequent drilling operations disturbed the quiet-time coherence results. 
\begin{figure}[t]
\includegraphics[width=8cm]{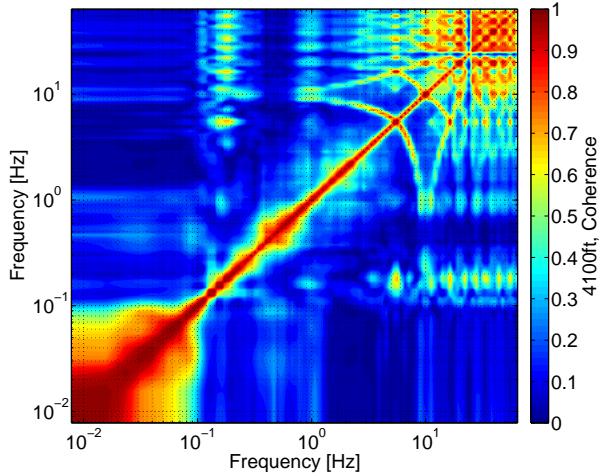}
\caption{The bifrequency plot shows the coherence between variations of quiet-time spectral densities at different frequencies. Gaussian stationary noise would yield a diagonal with coherence 1, and 0 elsewhere. Bifrequency coherence below 0.1\,Hz is due to earthquake signals. The variations of oceanic microseisms between 0.1\,Hz and 1\,Hz are only weakly correlated, which could mean that many different ocean-wave fields contribute simultaneously to oceanic microseisms. The patterns around 5\,Hz and 10\,Hz are caused by wandering ventilation lines. There is no conclusive
explanation for the structures above 10\,Hz.}
\label{fig:Bifreq}
\end{figure}
In Figure \ref{fig:Bifreq}, the bifrequency coherence calculated from quiet-time spectral densities is plotted. The (quiet-time) seismic field approximates a stationary Gaussian process only between 0.1\,Hz –- 4\,Hz. Earthquakes lead to correlation at frequencies below 0.1\,Hz. Ventilation lines cause curved coherence patterns centered near 5\,Hz and its harmonics. The coherence patch above 10\,Hz is of unknown origin. They could be the result of a Gutenberg-Richter type law \cite{GuRi1944}, i.e.~rates of events with different corner frequencies are linked to each other, or they could be caused by coupling to a broad-band acoustical source.
\begin{figure}[t]
\includegraphics[width=8cm]{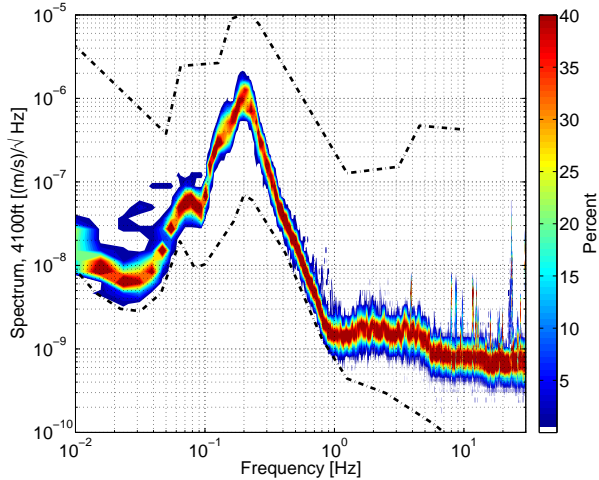}
\caption{The root spectral density of the seismic field at the 4100\,ft level is represented as spectral variation over one day (January 31st). Variations at all frequencies are very small. This is partially the result of the cleaning algorithm that removes the power of short transient events from the data. However, as is shown in Section \ref{sec:Blasts}, seismic events occur rarely during a day, and therefore the main effect of the cleaning algorithm is to reduce the width of the spectral distribution without changing the mean spectral density significantly. The two dashed curves correspond to the Peterson low- and high-noise models \cite{Pet1993}.}
\label{fig:Spec4100}
\end{figure}

We conclude this section with a comparison of average seismic spectra at different levels. The weakest seismic spectra are measured at the 4100\,ft level. The plot in Figure \ref{fig:Spec4100} shows the distribution of root spectral densities for each frequency over one complete day based on half-hour averages. The data were acquired on January 31st. In general, variations at the 4100\,ft level are small, consistent with our previous results. The dashed lines indicate the Peterson low and high-noise models \cite{Pet1993}. The Homestake spectrum at 4100\,ft depth around 1\,Hz lies only a small factor above the low-noise model. During summer when wind speeds on the oceans is generally smaller on the northern hemisphere, the low-noise model would be reached and even beaten around the microseismic peaks (50\,mHz -- 0.7\,Hz), probably thanks to the great distance of this site from all oceans. The plot is based on quiet-time data. The events removed from the data only increase the variations above 2\,Hz without affecting the mean value of the spectrum significantly (this is true for normal times like January 31st, while exceptional events can be sufficiently powerful to change the mean spectral density). The spectrum proves that the 4100\,ft level can provide an excellent low-noise environment. 

\begin{figure}[t]
\includegraphics[width=8cm]{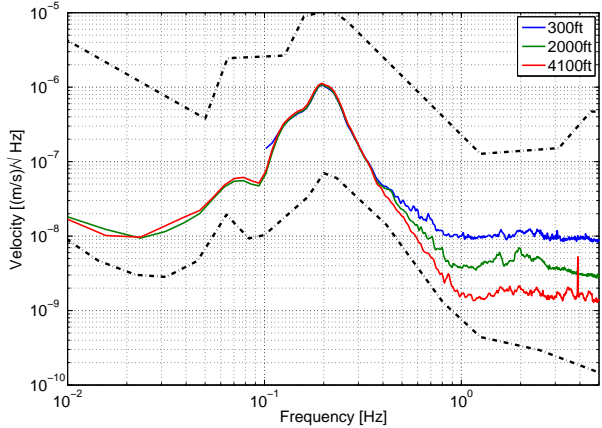}
\caption{The plot shows a comparison of seismic-noise spectra measured at 300\,ft, 2000\,ft and 4100\,ft depths. Seismic noise above 1\,Hz is about a factor 10 weaker at the 4100\,ft than at the 300\,ft level. The two dashed curves correspond to the Peterson low- and high-noise models \cite{Pet1993}. The spectra are drawn at frequencies corresponding to the pass-bands of the individual seismometers.}
\label{fig:SpecComp}
\end{figure}

The comparison of seismic-noise spectra is shown in Figure \ref{fig:SpecComp} for January 31st, 2010.
This day was not particularly quiet except for the fact that no teleseismic event occurred. The spectrum below 1\,Hz depends on seasonal variations. At 2000\,ft and 4100\,ft, ground motion between 2\,Hz--4\,Hz was stronger on January 31st than at most other days. The seismic noise above 1\,Hz decreases with increasing depth. It is about a factor 10 weaker in amplitude at the 4100\,ft level than at the 300\,ft level. This observation is consistent with previously published results \cite{Bor2002,CarEA1991,Dou1964}, and it constitutes the first advantage of underground GW detectors upon surface detectors. However, whereas the gain in terms of seismic noise coupling to the test mass through the suspension system is immediate, the benefit in terms of Newtonian noise generated by the seismic field is not easily evaluated. Here, multiple aspects of the problem have to be taken into account including the homogeneity of the rock and the number density of local seismic sources \cite{HaEA2009b}.

\section{Oceanic Microseisms}
\label{sec:Secondary}
Microseisms between 50\,mHz--0.3\,Hz are known to be generated by ocean waves. These frequencies comprise the primary microseismic peak, which is usually found below 0.1\,Hz, and the secondary microseismic peak at frequencies above 0.1\,Hz. The two peaks are related to two different source mechanisms. The primary peak is caused by ocean waves exerting pressure on the ground in shallow waters near the coast. As explained in \cite{LH1950}, the physics behind the secondary peak are more complicated. Two counter-propagating wave fields need to form a standing ocean wave that generates pressure oscillations at twice the ocean-wave frequency. Whereas pressure oscillations caused by a travelling ocean wave would decrease exponentially in amplitude towards greater water depths, oscillations from a standing wave can propagate all the way down to the ocean bottom where they are converted into seismic waves.
\begin{figure}[t]
\includegraphics[width=8cm]{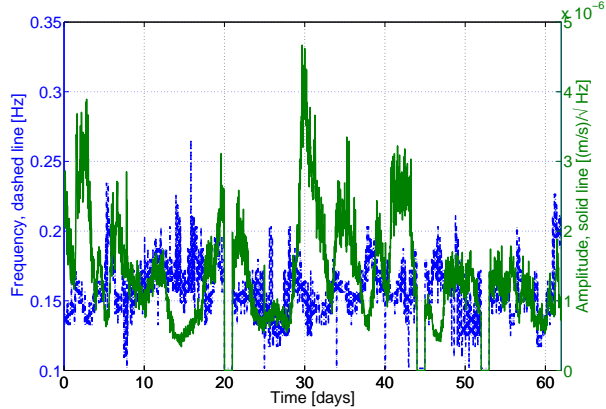}
\caption{The plot shows the frequency (dashed curve) and amplitude evolution (solid curve) of the secondary microseismic peak measured at the 4100\,ft level during December 2009 and January 2010. The ocean waves that generated the peak had twice the period of the seismic motion.}
\label{fig:Secmicro}
\end{figure}
In the past, sources of the microseismic peaks have been identified by beam forming using surface arrays. In \cite{FKK1998}, the authors determined the backazimuths of sources for the primary and secondary peak measured in Germany with the Gr\"afenberg array \cite{HaSe1978}. A mode-dependent analysis of microseisms was presented in \cite{ToLa1968} using the data from the LASA array in Montana, USA \cite{GFR1965}. They found that oceanic microseisms either originate from the direction of the Pacific Ocean, or the Labrador Sea.

\begin{figure}[t]
\includegraphics[width=8cm]{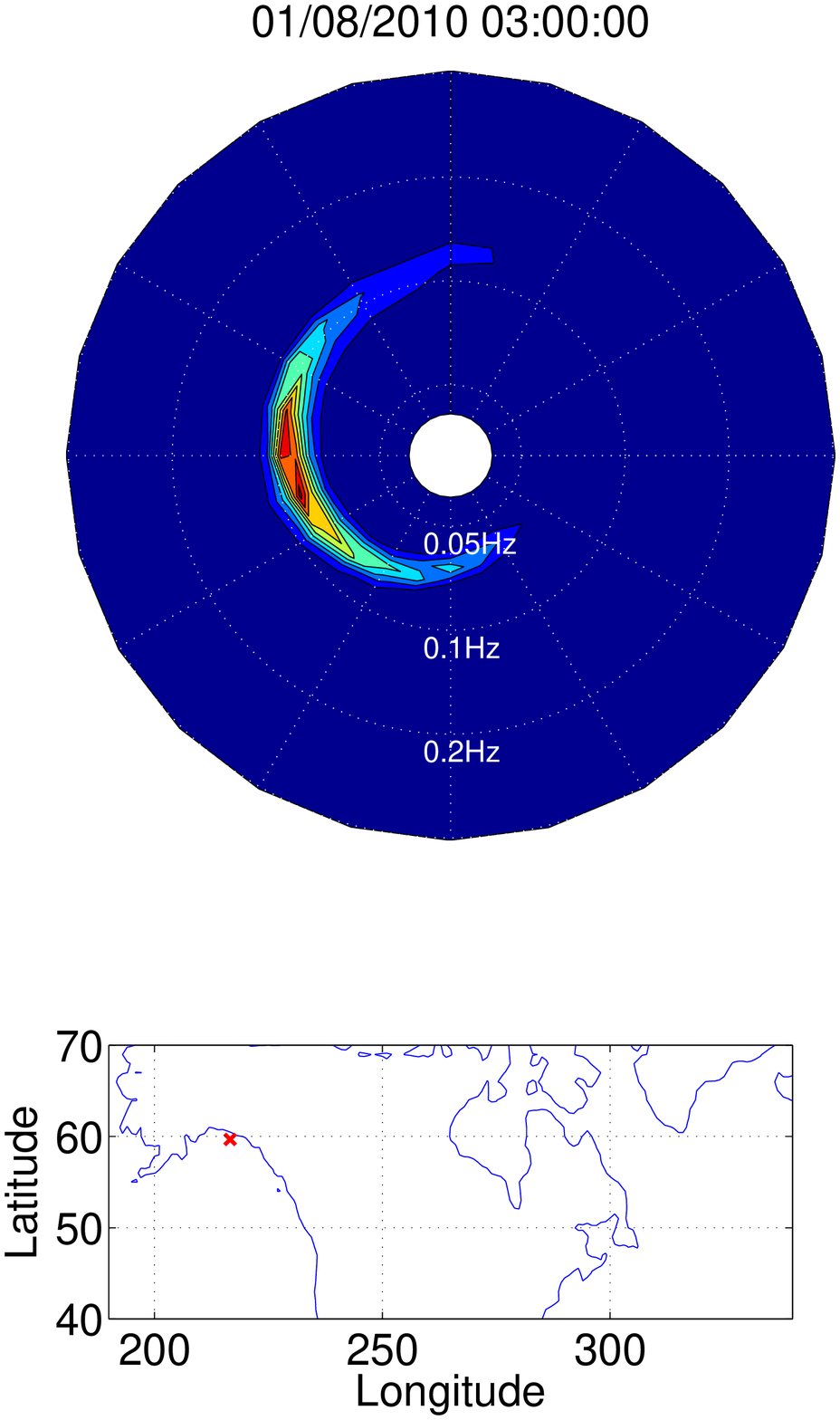}
\caption{The plot shows the wave spectrum at a buoy location near the coast of Alaska as a function of frequency and propagation direction. The data for this plot are a product of a model (NCEP wave model WAVEWATCH III \cite{Tol2009}) based on ocean wind data, which are validated at buoy locations comparing with measured wave heights. In this way, a propagation-direction dependent wave amplitude and frequency can be determined. The wave amplitude in this plot belongs to a swell whose propagation direction is spread over more than $180^\circ$, a typical shape at buoys close to Alaska. As pointed out before, the generation of secondary microseisms requires two counter-propagating wave fields of the same frequency. Here, a single field serves this condition due to its propagation spread. As will be shown in Figure \ref{fig:Buoy41}, wave frequency at this buoy location matched the frequency evolution of microseisms at Homestake very well over several days around January 8th.}
\label{fig:Buoyspec}
\end{figure}

In this section, we present a different approach to relate the secondary microseismic peak to its sources. As can be seen in Figure \ref{fig:Secmicro}, the frequency $f_{\rm sec}(t)$ and amplitude of the seismic peak vary significantly over time. Since ocean waves must have twice the period of the secondary microseisms, one can directly search the surrounding oceans for waves whose frequency evolution matches the evolution of the observed secondary peak. For this purpose, buoy data were studied that were downloaded from a server of the National Oceanic and Atmospheric Administration (NOAA) \cite{NOAA2010}. Within a geographic window around the US defined by the $40^\circ$ and $70^\circ$ northern latitudes, and the $20^\circ$ and $170^\circ$ western longitudes, data associated with 65 buoys are provided (a complete file contains buoys that are located all over the world). The type of data that can be extracted from the buoy files is displayed in Figure \ref{fig:Buoyspec}. 
It contains the frequencies, amplitudes and propagation directions of wave fields provided by the National Centers for Environmental Prediction (NCEP) wave model WAVEWATCH III \cite{Tol2009}. One ocean wave spectrum is produced every 3 hours for each buoy. These wave predictions are validated at buoy locations by comparison with measured wave heights. Ocean waves can be classified as swell or wind waves. Wind waves generated at high frequencies evolve into swell as soon as the waves separate from the wind that generated the waves \cite{Yo1999}. Swell is stabilized by non-linear processes and eventually acquires its typical oscillation period around 15\,s. Taking the frequency-doubling into account, this means that the secondary peak is produced by swell, not by wind waves. The wave field represented in Figure \ref{fig:Buoyspec} belongs to a swell. Its propagation spread exceeds $180^\circ$, which means that a significant part of the wave field forms a standing wave.  To search for waves with specific frequencies, we first identify the maximum wave amplitude between 0.05\,Hz--0.15\,Hz for all buoys and collect their frequencies $f_{\rm ow}(t)$. The next step is to define a method to compare the frequency $f_{\rm sec}(t)/2$ with the ocean-wave frequency $f_{\rm ow}(t)$. The main problem here is that throughout a month microseisms are generated at different locations on the ocean. A study that compares $f_{\rm ow}(t)$ at one fixed location with $f_{\rm sec}(t)/2$ during an entire month does not have to yield a good overall agreement between frequencies. Therefore, results from the long-term study of $f_{\rm ow}(t)$ at fixed buoy locations versus $f_{\rm sec}(t)/2$ will be compared, at least qualitatively, with short-term matches of these two frequencies. 

\begin{figure}[t]
\includegraphics[width=8cm]{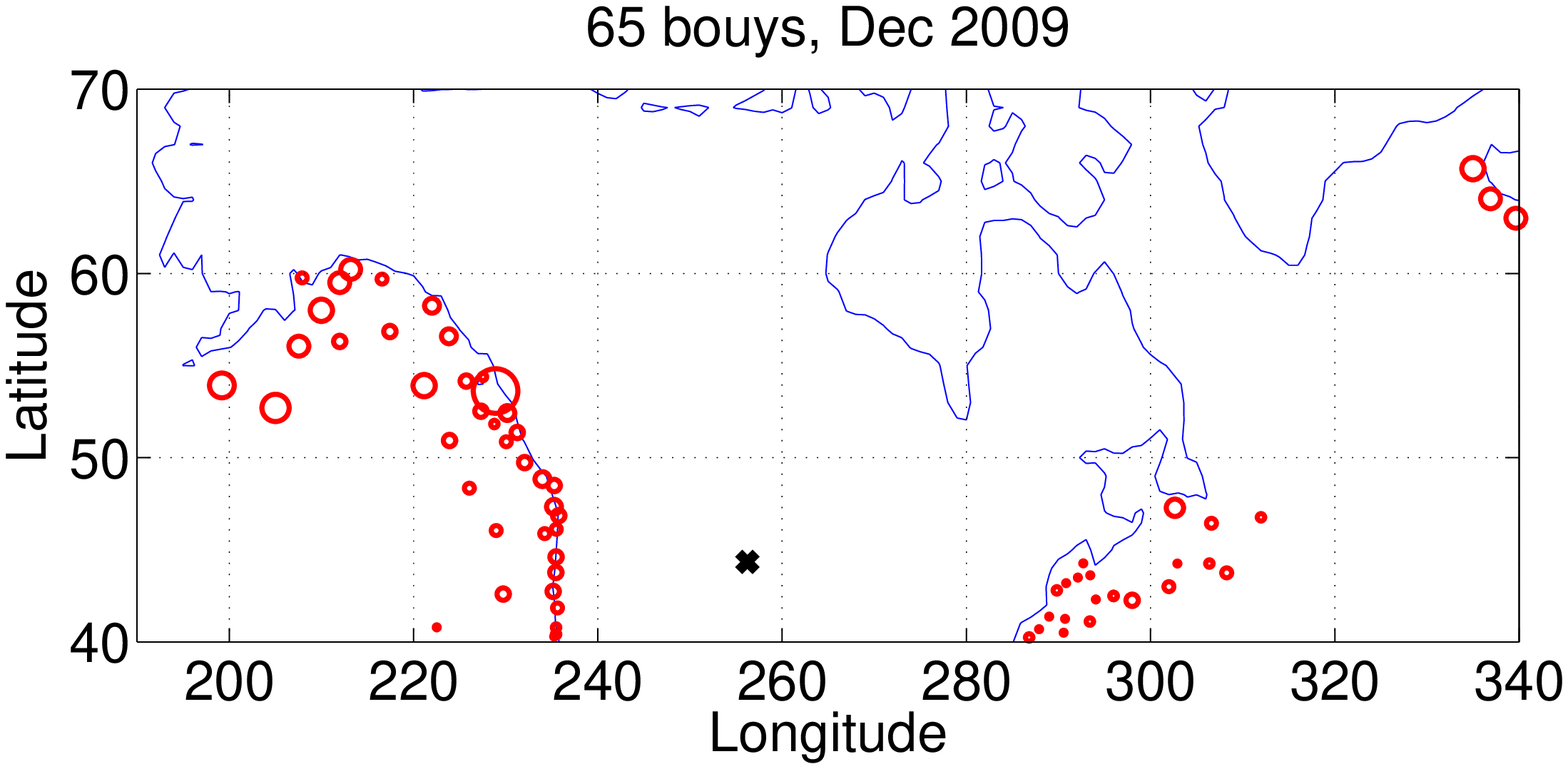}\\
\includegraphics[width=8cm]{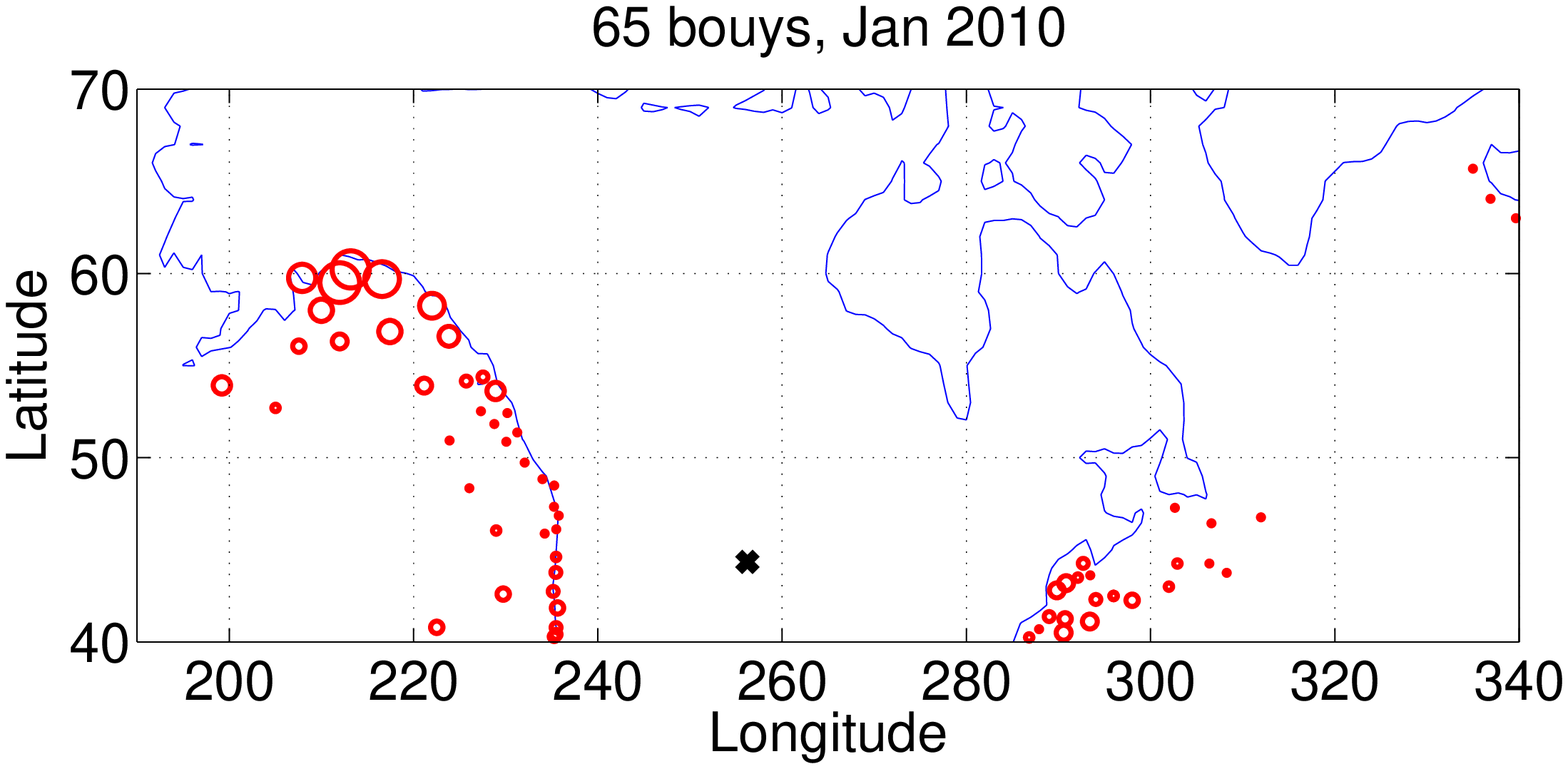}
\caption{The two plots show the long-term agreement between frequencies of ocean waves and microseisms measured at Homestake (marked with a cross). Each circle centered at one of the 65 buoy locations is drawn with a radius that is proportional to the coherence between the frequency evolutions divided by their mean-square deviation. In this way, the radii depend on absolute frequency values and on changes in frequency. The best agreement is found for buoys in the northern Pacific or northern Atlantic consistent with the results presented in \cite{ToLa1968}.}
\label{fig:Buoyscoh}
\end{figure}

\begin{figure}[t]
\includegraphics[width=8cm]{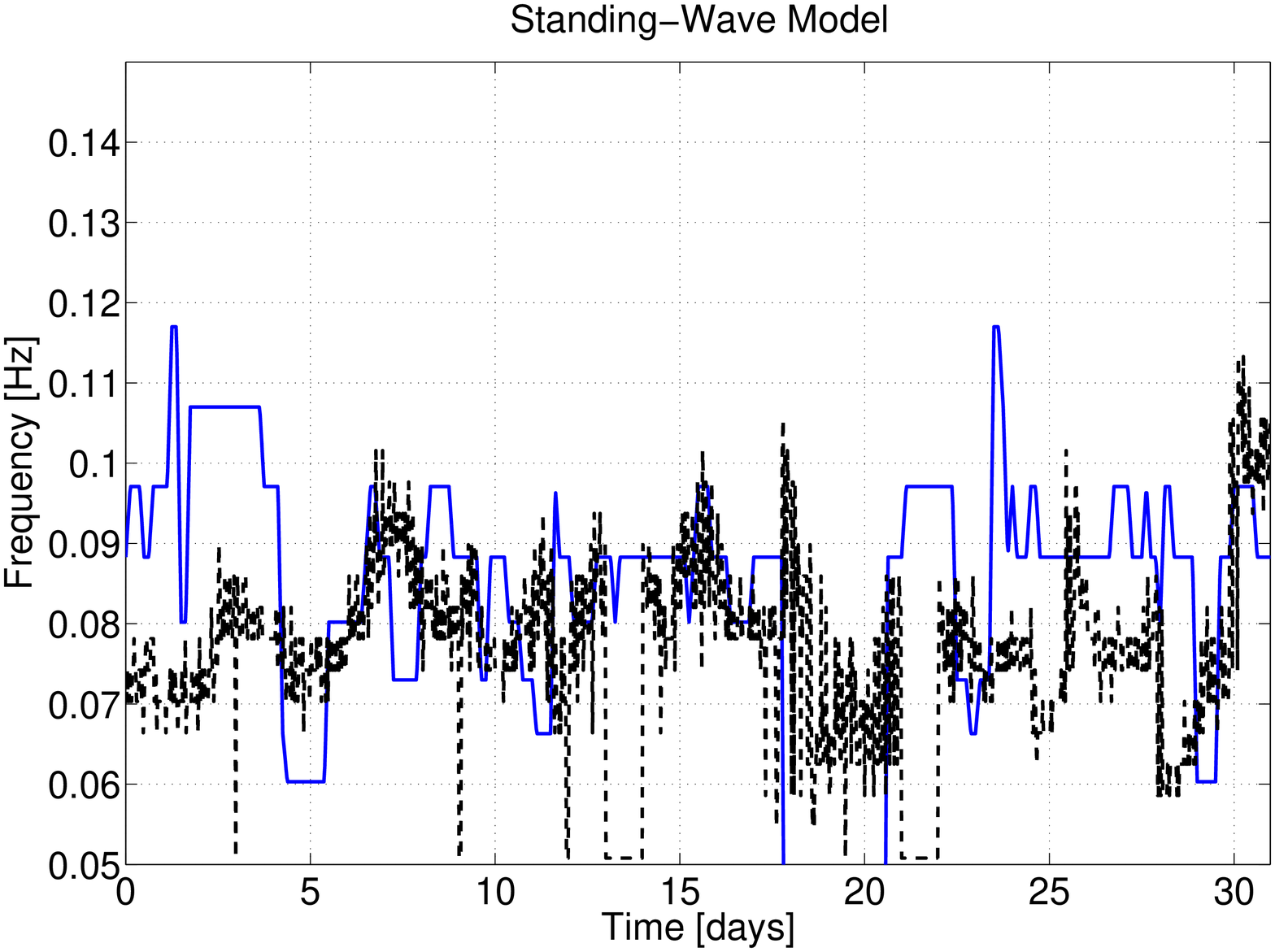}
\caption{The plot shows the evolution of the ocean-wave frequency $f_{\rm ow}(t)$, solid line, and of the secondary microseismic peak $f_{\rm sec}(t)/2$, dashed line, during January 2010. No other location has a better match between January 6th and January 24th, and after January 28th. This buoy also shows one of the best long-term matches.}
\label{fig:Buoy41}
\end{figure}

The results of the long-term analysis are shown in Figure \ref{fig:Buoyscoh} for December 2009 and January 2010. For each of the 65 buoys, a circle is drawn whose radius $R$ is proportional to the coherence between the frequency evolutions of ocean waves and microseisms divided by the mean-square deviation of the ocean-wave frequency and half the frequency of the secondary peak:
\beq
R\propto \frac{\sum\limits_i(f_{\rm sec}(t_i)-\langle f_{\rm sec}\rangle)(f_{\rm oc}(t_i)-\langle f_{\rm oc}\rangle))}{\sum\limits_i(f_{\rm sec}(t_i)/2-f_{\rm oc}(t_i))^2}
\eeq
The combination of these two measures guarantees that changes as well as absolute values of frequencies are compared. If the coherence is not taken into account, then the method would yield good matches with locations where wave frequency is almost constant around the mean value of $f_{\rm sec}(t)/2$. Results for both months identify the northern part of the Pacific as main source of microseisms measured at Homestake.
This is confirmed by further investigations of individual buoy locations. In Figure \ref{fig:Buoy41}, the ocean-wave frequency during January 2010 is shown for the same buoy as in Figure \ref{fig:Buoyspec} together with $f_{\rm sec}(t)/2$. Between January 6th and January 24th, and after January 28th, the frequencies match very well. The same buoy also has the second best long-term match in January. Buoys with poor long-term match also showed no significant short-term match. 

As was pointed out before, one characteristic of the wave field at buoys with good matches in January was that the buoy measured a single swell with a large propagation spread so that the ocean-wave field could form a standing wave by itself. The wave fields in December at buoys with good matches had the same shape most of the time. Occasionally, a few distinct counter-propagating fields were measured at these locations (we need to keep in mind that these are modelled waves consistent with measured wave heights). The fact that wave fields at all buoys with good matches show a single swell with large spread or two counter-propagating swells is further evidence for the generation model of secondary microseisms. The spectra for the majority of buoys with a bad match either show several swells at different frequencies or one swell with a small spread, in both cases the swell could not produce a standing wave. 

The results presented in this section indicate that microseisms at a specific frequency (peak frequency) can be associated with wave fields at specific locations, and that these locations change over the course of a month. The best matches of frequency evolution were found at near-coast buoys, which suggests that standing-wave fields are more likely to occur near the coast. It may also be that conversion of oceanic pressure fluctuations into seismic waves is more efficient above the comparatively shallow continental shelves. However, due to the lack of buoys at high sea, this conclusion may also follow from an observation-selection effect.

\section{Conclusion}
\label{sec:Conclude}
Our analysis shows that the deeper levels of the former Homestake mine provide a formidable low seismic-noise environment. The average seismic-noise spectra approach the global low-noise model  over a considerable fraction of the observation time. Seismic spectral densities vary weakly over the course of the two months of observation. Most of the spectral variation comes from short seismic events. Event rates at Homestake are small despite the regular blasting and possible rock fracturing as a response to blasting and dewatering. The vast majority of events occurs during the working hours between 6am and 4pm local time. The sources of these events (either underground or at the surface) need to be located to determine whether they could be avoided in a GW observatory. However, the main factor to determine quiet times is the performance of the seismic noise or NN filters as a function of event magnitude.

Sources of the secondary microseismic peak were located in the northern Pacific during the two months December 2009 and January 2010. The exact locations of the standing-wave fields were shown to be changing over time. The most likely origins of secondary microseisms were identified nearer to the coast, which could mean that the generation of these microseisms is more efficient on the continental shelf. We found that these wave fields at times when the ocean-wave frequency matched (half) the frequency of the microseisms were able to form standing waves, either by self-interference or occasionally by interference of two counter-propagating fields that had similar frequencies. The ocean-wave model that produces the ocean-wave spectra at buoy locations can in principle output a wave field of the entire Pacific Ocean (and any other ocean). It would be a significant improvement if the search for the sources of microseisms was not constrained to the buoy locations. However, a global ocean-wave model is currently unavailable. 

At frequencies below the microseismic peaks, we found that part of the seismicity at the 300\,ft station is related to surface wind speeds and unaffected at higher depths. The fact that surface wind speed has no influence on seismicity at deeper levels suggests that the wind-generated seismic surface fields are characterized by much smaller length-scales than one would expect assuming Rayleigh-wave speeds of a few km/s (which is typical for hard rock). A possible explanation is that the wind-generated surface waves propagate within a low-speed surface layer and decay in amplitude over a much shorter vertical distance corresponding to the thickness of the layer. One interesting future experiment would be to deploy several barometers and anemometers at the surface that can synchronously sample pressure and wind speed at higher frequencies, and to study their correlations with seismic fields as a function of depth. In addition, the problem should be studied for other sources like surface pressure fluctuations or even anthropogenic sources. These studies would have a great impact on Newtonian noise models. It is still uncertain to what level surface effects need to be incorporated into the models of underground Newtonian noise. Our results prove that a simple understanding in terms of individual evanescent fundamental Rayleigh waves does not lead to accurate predictions of underground seismicity, and therefore underground Newtonian noise.

Many important problems relevant to Newtonian-noise modelling cannot be investigated with the current seismic array, even factoring in the forthcoming improvements of the readout system and synchronization of the array. First, a denser and wider seismic array extending in 3D is required to provide a high-quality spatial spectrum of the seismic field. Newtonian-noise models depend on the nature of seismic sources, whether they are distant or local, at the surface or underground. The current array does not have sufficient beam forming capabilities, nor does it allow us to study the mode content of the seismic field. All seismometers happen to lie approximately within a vertical plane (other seismometer locations were not available at the time of installation), and the array consists of too few seismometers to provide good spatial resolution and suppression of spatial aliasing. Using simple correlation methods to calculate the spatial spectrum, the required number of seismometers would be in the thousands. However, it should be clear that excellent seismic instruments and readout systems would facilitate the use of SNR-based techniques like maximum-likelihood methods, which require a much smaller number of seismometers. It is still an open question though how to estimate the required number of seismometers as a function of SNR (seismic noise relative to instrumental noise). This is one of the important problems that need to be solved in the future. 

\section{Acknowledgments}
We thank the staff at the Sanford Underground Laboratory and the LIGO Laboratory for their support. LIGO is operated for the National Science Foundation (NSF) by the California Institute of Technology under Cooperative Agreement PHY-0757058. This work has been performed with the support of the European Commission under the Framework Programme 7 (FP7) Capacities, project Einstein Telescope design study (Grant Agreement 211743), http://www.et-gw.eu/. This work is part of the research programme of the Foundation for Fundamental Research on Matter (FOM), which is financially supported by the Netherlands Organisation for Scientific Research (NWO). The work of JH, SD, SK, and VM was supported by the NSF Grant No.~PHY-0758036 and by the University of Minnesota. NC's work was supported by the NSF Grant No.~PHY-0854790, and Carleton College. Our summer students, AS, LN and TOK, gratefully acknowledge the support from INFN in the framework of the Virgo/LIGO undergraduate student exchange, and from NSF-LIGO in the framework of the Caltech SURF program. We thank Larry Stetler from SDSMT for providing the data from the Homestake weather station, Arun Chawla from NOAA for assisting with the ocean-wave data, and Warren Johnson and Martin McHugh for the loan of the STS-2.

\raggedright
\bibliography{c:/MyStuff/references}

\end{document}